# Influence of random roughness on the adhesion between metal surfaces due to capillary condensation

P.J. van Zwol, G. Palasantzas[*] and J. Th. M. De Hosson

Department of Applied Physics, Netherlands Institute for Metals Research and Zernike Institute for Advanced Materials, University of Groningen, Nijenborgh 4, 9747 AG Groningen, the Netherlands.

**Abstract**

The capillary force was measured by atomic force microscopy between a gold coated sphere mounted on a cantilever and gold surfaces with different roughness. For smooth surfaces the capillary adhesive force surpasses in magnitude any dispersion, e.g. van der Waals/Casimir, and/or electrostatic forces. A substantial decrease in the capillary force was observed by increasing the roughness ampltitude a few nanometers between *1-10 nm*. From these measurements two limits can be defined: a smooth limit where a closely macroscopic size contact surface interacts through the capillary force, and the rough limit where only a few asperities give a capillary contribution.

*Pacs numbers:* 85.85.+j, 68.55.-a, 47.55.nb, 68.03.-g, 07.10.Cm

e-mail: g.palasantzas@rug.nl

Adhesion is an important phenomenon in emerging technologies related to microelectromechanical systems (MEMS), where preventing sticking of device components is a major challenge. In general MEMS are sensitive to adhesion as a result of the large surface-to-volume ratio, of the micro- to submicrometer separations, and due to the highly compliant components. Dispersive surface forces (e.g. van der Waals/Casimir), which are relatively well understood [1, 2], contribute to adhesion and play a major role during the fabrication and operation of MEMS [3–8]. However, complications arise when dealing with real systems due to surface roughness and capillarity driven forces.

Capillary condensation occurs into small cracks and pores by wetting of liquids that originate from water vapor. As a result when the planar surfaces of MEM microswitches come in close proximity attractive capillary forces can start acting on them. The latter occurs if the surface separation becomes $\approx 2R_k cos(\theta)$ with $\theta$ the contact angle.[9] At equilibrium, the Kelvin radius $R_k$ (which is related to the meniscus height) is related to the relative vapor pressure $P/P_{sat}$ at temperature $T$ by the Kelvin equation $R_k = -(\gamma V_m/RT)[\log(P/P_{sat})]^{-1}$ [1, 9]. $\gamma$ is the liquid vapor pressure (e.g. $\gamma=73\ mJ/m^2$ for water), and $V_m$ is the molar volume of the liquid. However, the process by which asperity phenomena bridge the length scale causing adhesion of micro/macroscopic interfaces, is still unknown.

Recently, it was shown that below a threshold of relative humidity (RH) $P/P_{sat}$ the adhesion was mainly due to van der Waals forces across extensive non-contacting areas between Si-oxide surfaces, while above this threshold of RH, capillary condensation was the predominant term [10]. The adhesion energy was found to vary from $10^{-3}$ to $10^2\ mJ/m^2$, with values highly depending on the surface roughness (the plate roughness varied from *2 - 10 nm,* while the cantilever roughness was not mentioned) and high RH (threshold *≥70 %* depending on roughness). In former studies of the roughness influence on adhesive forces, either surfaces with a relatively high roughness or a low amount of samples were used [11].

Therefore, in this paper, we will investigate the adhesion force compared to other dispersion forces (van der Waals/Casimir) between gold (Au) surfaces as a function of the nanoscale roughness amplitude at a relative humidity below *70 %*, and the lateral correlation length. For this purpose the roughness of the plate was systematically and carefully increased from atomically flat annealed films on mica up to approximately *10 nm* rms roughness amplitude for vapor deposited Au films.

The Picoforce AFM from Veeco was used for the force measurements, where creep and hysteresis are eliminated with the closed loop scanner (having noise level below *<0.5 nm* and a nonlinearity of *0.1%* over a *20 μm* range). Adhesive forces were measured by retracting a sphere attached on the cantilever from the surface. The spring constant is measured from *8* electrostatic force curves with applied voltages between *±3 V* to *±4.5 V* (error *1 mV*), at a distance above *1 μm*, yielding *k=5 N/m* (error less *<10 %)*. Polysterene spheres (*100 μm* in diameter, $2R_s$, with 1.5% deviation from sphericity) were used. AFM was employed to determine the roughness of the sphere and yielded *1.2 nm* rms roughness amplitude (before coating) over a *25 μm²* area. The spheres were coated with *100 nm* Au in an evaporator (at *10⁻⁶ mbar* base pressure) and at a deposition rate *0.6 nm/sec* yielding surfaces with *1.8 nm* rms roughness. The force measurements were repeated with different spheres having the same rms roughness amplitude and cantilevers. After the force measurement, inspection with SEM did not reveal any damage to the Au layer on the sphere. In addition, successive AFM scans revealed the typical roughness of the Au film on the sphere indicating that the Au layer remained unmodified by the measurement process. The contact potential has been eliminated within *20 mV* error, which resulted in electrostatic forces much smaller than the van der Waals force at closest separation.

In this setup, the sphere surface sets a lower limit on the combined smoothness of the adhering objects. Therefore it is important to investigate whether the perfectly smooth limit is reached. Au films from *100* up to *1200 nm* were grown on silicon wafers and *150 nm* on mica. The latter sample was annealed extensively to obtain very smooth surfaces (see Fig. 2). The measured roughness varied from *0.8 nm* rms for the Au/mica and up to *1.1-9 nm* for the Au/SiO$_2$ films. Note however that the Au/mica film is very smooth locally, indicated by the large correlation length $\xi \approx$ *100 nm* and its height variations of atomic steps in size. The correlation length of the Au/SiO$_2$ films is much smaller within $\xi \approx$ *15-40 nm*. For gauging the effect of correlation length on the adhesive force, another film (*120 nm* thick) was made on a smoother wafer, which resulted in an rms roughness of *1.5 nm* but a correlation length about three times larger than that of the *100 nm* thick Au/SiO$_2$ film (Fig 2).

Figure 3 shows that the adhesion force remains very large for weakly rough surfaces, and follows a rapid decrease with more than two orders of magnitude for a variation of the rms roughness amplitude by a few nanometers in the range *1-10 nm*. This range of roughness is typical for fabricated NEMS surfaces [10], and therefore sets the limit over which capilary forces are of importance in nanodevices covered with metallic layers. The results for the *120 nm* film (*1.5 nm* rms roughness) with large correlation length of *60 nm* did not yield higher forces than the *100 nm* thick film with a correlation length of *20 nm* (also *1.5 nm* rms). Therefore, we can conclude that the rms roughness gives the predominant effect on the adhesive force due to capillary condensation.

From the roughness scans on different locations on the surface, the variation in the RMS roughness is ~*0.1 nm*, while the standard deviation in adhesion force was *30-70%* (Fig. 3) for the rougher films when measuring at *50-100* different surface locations. The latter is larger than the error due to calibration of the cantilever spring constant, and originates from a set of different reasons. First it is important to realize that the capillary force is extremely sensitive to nanoscale roughness, where it falls two orders of magnitude with increasing rms roughness in the range *1-10 nm* (Fig. 3). Therefore, it is not surprising of obtaining a force a few times smaller on the same plate if there are some local protruding peaks that may also deform. The latter was visible in some films where the force gradually increased up to five times after immediate successive measurement at the same location. All of these effects were, however, not seen for the Au/mica film, which is locally much smoother and does not have any peaks that deform easily. In contrast high peaks were abundant in the other Au films and also the correlation length was found to be small. The load was kept low during approach to avoid significant influence on the adhesion force beyond the stated error bars. Another explanation for the force deviation originates from STM studies, which indicated that below 60% RH the water film on Au is not closed but droplet formation takes place (even for RH as low as 15%) [12]. The average height of droplets *~5-10 nm* did not decrease much with lower RH, while the amount and size of droplets was decreased [12].

The total adhesion force can be divided into a capillary force and an interfacial tension force due to surface tension acting tangentially to the interface along the contact line with the solid body [11]. For the relatively smooth films, the theory prediction for flat surfaces gives an approximate upper limit for the force $F_{up} = 4\pi\gamma R_s \cos(\theta)$ [11] ($F_{up} = S\Delta\sigma$ with $S = 4\pi R_s R_k \cos(\theta)$ the meniscus surface area and $\Delta\sigma = \gamma/R_k$ the Laplace pressure, while ignoring contributions from surface tension). For the contact angle of water onto Au surfaces of *70°* [12] we obtain for $R_s$=*50 μm*, $F_{up}$=*1.5x10$^4$ nN*. With increasing roughness, surfaces can be more hydrophobic [13] leading to reduced capilary force. In addition, the contact area between sphere and plate is highly reduced. If we consider the lower limit for the force in terms of contact onto a single asperity [14], where the capillary meniscus is formed only between this asperity and the surface of the sphere, we obtain a force of magnitude $F_{low} = 4\pi\gamma\xi\cos(\theta)$ =*15 nN* (where we used as an effective asperity radious the roughness correlation length *ξ)*. From Fig. 3 it appears that the smooth limit is reached for the Au/mica film. For the roughest films the values found are up to ten times higher than that of a single asperity indicating a capillary interaction of a multitude of asperities.

When the RH is changed from 15% up to 60% no significant change in capillary adhesion was observed, although we do not rule out any change because the standard deviation in the measurement is rather large. Again the STM analysis of the incomplete wetting of the water layer mentioned earlier give a good explanation for this behavior [11]. In [10] it was reported that for very high RH close to *100%*, corresponding to large kelvin condensation length, the capillary force reaches a maximum value even for the roughest surfaces. This is not surprising as at high RH the water layer on the film increases rapidly even surpassing the highest roughness peaks. The capillary force overshadows any other force (van der Waals/Casimir or electrostatic) as being *100-1000* times stronger for smooth surfaces. For rough surfaces the Casimir force becomes comparable within an order of magnitude to the capillary force. The Casimir force between plate and sphere is roughly *5 nN* for the rough surfaces in the range of contact, whereas similar sized capillary forces (*15 nN*) are produced by single asperities with a lateral size of about *30 nm*.

In conclusion, we have shown that for relatively low humidity capillary forces are present in the case of smooth surfaces, and surpasses in magnitude any dispersion and electrostatic forces. In addition, an enormous decrease in the capilary force was observed by increasing the roughness ampltitude a few nanometers in the range *~1-10 nm*. Considering the rapid fall off in the capillary force and the two limits (a smooth limit where the whole surface contribute to the capillary force, and a rough limit where only a single or a few asperities contribute), the crossover regime might in turn depend on the contact angle and any latteral roughness features. Both could be an interresting direction for further study of this phenomenon in the design MEMS/NEMS if stiction poses a problem.

**Acknowledgements**: The research was carried out under project number MC3.05242 in the framework of the Strategic Research programme of the Netherlands Institute for Metals Research (NIMR). Financial support from the NIMR is gratefully acknowledged. We would like to thank H. Akkerman, J. Harkema, and S. Bakker for providing us with Au Samples.

**Figure captions**

**Figure 1**: Surface scans of the *100nm*, gold on mica, *280 nm* films and the coated sphere surface, the scan size is *2μm* and the color range corresponds to *25 nm* z-range, where lighter shades mean higher areas.

**Figure 2** Plots of the roughness amplitude w versus lateral correlation length ξ for the gold grown on *SiO$_2$*, Mica and the sphere.

**Figure 3** Adhesion force vs. rms roughness of the sphere an plate added together. The upper and lower horizontal lines represent the theoretical values for capillary interaction, in case of a perfectly smooth sphere with a perfectly smooth plate, and in case of a single aspherity with correlation length *50 nm*.